\documentstyle[prl,aps]{revtex}
\input{psfig}
\draft 
\begin{document}
\twocolumn[\hsize\textwidth\columnwidth\hsize\csname@twocolumnfalse\endcsname
\title{Electrostatically-Driven Granular Media: Phase Transitions 
and Coarsening }
\author{I.S. Aranson, D. Blair, V.A.  Kalatsky, 
G.W. Crabtree, W.-K. Kwok, V.M. Vinokur and  U. Welp}
\address{Argonne National Laboratory, 9700 South Cass Avenue, Argonne, IL 60439}
\date{\today}
\maketitle

\begin{abstract}
We report the  experimental and theoretical study of
electrostatically driven granular material.
We show that the charged granular medium
undergoes a hysteretic first order phase transition from the immobile
condensed state (granular solid)
to a fluidized dilated  state
(granular gas) with a changing applied electric field.
In addition we observe a
spontaneous precipitation of  dense
clusters from the gas phase and subsequent coarsening -- coagulation of
these clusters.
Molecular dynamics simulations shows qualitative agreement with
experimental results.
\end{abstract}


\narrowtext
\vskip1pc]

Despite extensive study over the preceding decade,
a fundamental understanding of the dynamics of
granular materials still poses
a challenge for
physicists and engineers \cite{jnb,kadanoff}.  Peculiar properties of 
granular materials can be attributed to strong contact
interactions and inelastic collisions
between grains  
\cite{swin3,argonne}.  Fascinating collective behavior appears
when small particles acquire an electric charge and respond to 
competing long-range electromagnetic 
and short range contact forces.

The electrostatic excitation   of 
granular media  offers unique new 
opportunities compared
to traditional  vibration techniques which have been developed 
to explore granular dynamics \cite{swin3,argonne,textbook,beringer}. 
It enables one to deal with
extremely fine powders which are not easily controlled
by mechanical methods. 
Fine particles are  more sensitive
to electrostatic forces which arise through 
particle friction or space charges in the particle environment.
Their large surface to
volume ratio amplifies the effect of water or other surfactants.  
These effects intervene in the dynamics causing agglomeration, 
charging, etc., making mechanical
experiments uncontrollable. 
Electrostatic driving makes use of these {\it bulk} forces, and  allows 
not only for removal of these "side effects,"
but also for control by long-range electric forces. 

In this Letter we report experimental and theoretical   study of
electrostatically driven granular material.
It is  shown that the charged granular medium
undergoes a first order phase transition from the immobile
condensed state (granular solid)
to a fluidized dilated  state
(granular gas) with a changing applied electric field.
A spontaneous precipitation of  dense
clusters from the gas phase and subsequent coarsening -- coagulation of
these clusters is observed in a certain region of the electric field values.
We find that the  
rate of coarsening is controlled by the amplitude and frequency of the
applied electric field. We have also performed molecular dynamics simulations of
electrostatically driven particles. These simulations show
qualitative agreement with experiments.

The experimental cell is shown in Fig. \ref{Fig1}. 
Conducting particles are placed between 
the plates of a large capacitor which is energized by 
a constant or alternating  electric field. 
To provide optical access to  the cell, the upper  conducting 
plate is made transparent. We used $4\times6$ cm capacitor plates 
with a spacing of 1.5 mm and 35 $\mu$m copper powder.
The field amplitude varied from 0 to 10 kV/cm
and the frequencies on the interval of 0 to 250 Hz. 
The number of particles in the 
cell  was about $10^7$.
The experiments were performed both in atmospheric pressure and in 
vacuum ( $5\times 10^{-6}$ Torr).

When the conducting particles are in contact with the 
capacitor plate they acquire a surface  
charge. As the magnitude of the electric field 
in the capacitor exceeds the 
critical value $E_1$ the resulting (upward) electric 
force overcomes the gravitational force $mg$ ($m$ is the mass of the particle, 
$g$ is the acceleration due to gravity) 
and pushes  the charged particles upward.   
When the grains hit the upper plate, they deposit  their charge and fall 
back. 
%when this process repeats a gas-like phase is present. 
Applying an alternating electric field
$E=E_0 \sin (2 \pi f  t)$, and adjusting its frequency $f$
one can control the particle elevation by effectively turning them back  
before they collide with the upper plate.

%Charging of the particles results in electric current. 
%For the value of $E$ about $2E_1$
%it achieves   $1$ $\mu$a. 
The phase diagram is  shown in Fig. \ref{Fig2}. 
%Isolated particles start to move at $E>E_1$. 
We have found that at amplitudes of the electric field 
above a {\it second} threshold value, $E_2>E_1$, the granular medium 
forms a uniform gas-like phase (granular gas). 
This second field $E_2$ is 50-70\% larger  then  $E_1$ in nearly the
whole range of the parameters used. In the field interval $E_1<E<E_2$,
a remarkable phenomenon analogous to coalescence dynamics 
in systems exhibiting first order phase transitions \cite{coales,lp} was 
observed.  Upon decreasing the field below 
$E_2$, the gas phase loses 
its stability and small clusters of immobile  particles
surrounded by the granular gas   
form. 
These clusters then evolve via coarsening dynamics: 
small clusters disappear and large clusters grow. Eventually the 
clusters 
assume almost perfect circular form (Figs \ref{Fig3}a-c). 
In the process of coarsening the  electric current 
also changes in time due to decrease of particles 
in the  gas phase. 
A close-up image of one of the clusters is shown in Fig. \ref{Fig3}a. 
Surprisingly, we can even see a larger particle near the center of the cluster 
which served as the nucleation point at the early stages of the
aggregation. Due to its increased mass the large particle is the  {\it first} 
to become immobile in the lower field.  After a very large 
time ($t\approx 30000$ sec)  a single cluster containing about 
$10^6$ grains survives. 
At the final stage a dynamic equilibrium between the
granular solid and the surrounding gas 
persists -- not all the particles join the last cluster.

For the cell under atmospheric pressure (open cell)  
we found that both fields 
$E_{1}$ and $E_{2}$ grow as a  function of frequency for large $f$ and show 
non-monotonous behavior for $f\approx 12 $Hz. This indicates 
a characteristic time of the order $100$ msec. We suggest 
that cohesion may be responsible for this relatively large time. Indeed, 
due to the humidity of the air a surface coating should 
exist, requiring a characteristic time $\tau$ 
for the grain to detach from the capacitor plate. 
%Although the cohesive forces are very small, they become comparable
%with the gravitational/electric forces. 
In order to reduce the cohesion  we evacuated the cell
to $5\times 10^{-6}$ Torr. 
As demonstrated from Fig. \ref{Fig2}, the frequency dependence 
is indeed substantially reduced and becomes almost flat. Small 
oscillations in the dependence for low frequency are probably 
due to a  residual coating on the particles which
does not completely evaporate in vacuum. 

We have measured the number of clusters $N$ and the averaged
 radius of the clusters 
$\langle R \rangle$ as a function of time $t$ (see Fig. \ref{Fig4}) 
using images taken in time-lapse by a digital camera and then 
processed on a computer.
The values  of    $ \langle R \rangle$ for both open and evacuated cells 
are consistent with a power law  $ \langle R \rangle  \sim \sqrt t $.  
However, there is a substantial difference in the behavior of 
$N$. In open cells (no vacuum)  we observed  slow coarsening 
$N \sim 1/ \sqrt t$; whereas the coarsening  for the evacuated cell is much faster, 
$N \sim 1/t$. Remarkably, for the evacuated cell the exponents 
for $N$ and $ \langle R \rangle$ coincide with the exponents 
for two-dimensional bi-stable  systems in the interface-controlled 
regime of Ostwald Ripening  \cite{wagner,ams}. 
We speculate that   the slow-down in coarsening  for open cells 
can be  a result of 
cohesion between the particles and with the capacitor plate. 
This cohesion reduces the mobility of particles at the edges of the 
clusters, resulting in "pinning" of the front between the granular  solid and 
gas. 
%The effect of the interstitial air can be ruled out 
%because both critial fields $E_{1,2}$ show similar frequency dependence. 
%However, at the onset 
%of particle  motion, i.e. for $E_1$,  the displacements are 
%very small and the viscous drug 
%force is negligible.  

Let us 
compare the forces 
exerted on an isolated particle and on a particle 
which is held within a
cluster at the  same field.
The force between an isolated sphere and the capacitor plate in contact can be 
found as a limit of the problem of two spheres when 
the radius of one sphere goes to infinity. This problem had been
contemplated by several outstanding physicists  
including Kelvin, Poisson, Liouville and Kirchhoff \cite{sphere}.   
Building on the technique of Ref.  \cite{sphere}, we arrive at 
the force $F_e$ in question: 
\begin{equation} 
F_e=  c a^2 E^2
\label{eq1}
\end{equation} 
where $a$ is the radius of the sphere and $E$ is the field in the capacitor. 
The constant $c=\zeta(3)+1/6 \approx 1.36$ comes from summing of infinite 
series derived in \cite{sphere}. 
The electric force $F_e$ has to counterbalance the gravitational 
force $G=4/3 \pi \rho g a^3$, where $\rho$ is the density of the material. 
Comparing $F_e$ and $G$ we find the first critical field $E_1$:
\begin{equation} 
E_1 = \sqrt{\frac{4 \pi \rho  g a}{3\cdot 1.36}}
\label{eq2}
\end{equation} 
Our theory indicates no frequency dependence of $E_1$. 
For the parameters of our experiments we obtain $E_1=2.05 $ kV/cm. 
The  experimental critical field is $E_1\approx 2.4 $ kV/cm. 
This discrepancy seems reasonable since we did not take into account  
additional molecular and/or contact forces which will
increase the critical field.

If the spheres are close to each other, 
the surface charge will redistribute: 
each sphere in the cluster 
acquires smaller   charge  
than that of an 
individual sphere
due to a screening of the 
field by its neighbors. 
The exact derivation of the force acting on the
particle within the cluster is not available at present. 
The upper bound can be obtained  by replacing 
the square lattice of spheres with radius $a$  by 
a lattice of squares with 
area $4a^2$. 
Since  the charge density  of  the corresponding 
flat layer is  $\sigma=E/4 \pi$,  
the total electric force on the 
square  is $F_{2}= 4 a^2 \sigma E/2$.  
The ratio of the fields to lift 
individual particle $E_1$ and the particle in the square lattice $E_2$ is 
$
E_2/E_1 = \sqrt {2 \pi \cdot 1.36 } = 2.92$.
For closed-packed hexagonal lattice one obtains a slightly  higher value
$E_2/E_1\approx 3.14$. 
As one sees, the ratio of the critical fields is independent of particle 
size and the density and the frequency $f$. 
%For the constant applied field $E$, and, correspondingly, 
%for the low frequency ac field, the ratio of the fields is about 3.
This is consistent with the experimental findings,
see Fig. \ref{Fig2},  although it exceeds the 
experimental value by a factor of 2. 
A more accurate account for the surface shape will improve the ratio.

We have also performed molecular dynamics 
simulations of conducting spheres in applied electric field. 
Several simplifications have been implemented:
the sphere polarization  was neglected;   
the charge was assumed to be in the center of the sphere;
all collisions were assumed to be inelastic. 
These simplifications are justified by the
fact that the particle-particle collisions are
rather rare in contrast to
deep vibrated granular layers studied in Ref. \cite{bizon}.
Interactions between the 
spheres   are treated as being 
between point charges if the distance is smaller
than the plate spacing.  For larger distances we assumed no interaction 
due to screening of the far  charge  field by other particles
(compare with Debye screening \cite{lp}).

In our molecular dynamics simulations 
we implemented explicitly that the charge acquired by each  sphere 
from the conducting plate decreases gradually 
with the number of nearest neighbors: 
it is maximum for an isolated sphere  and decreases
by factor of two for the sphere  inside the cluster. 
We simulated  $35$ $\mu$m diameter 3000 copper spheres 
in the domain $7\times 7$ mm ($200\times 200$ particle diameters)  
with periodic boundary conditions. The capacitor spacing was   
$1.5$ mm, as it was in the experiment. Although  the quantitative correspondence 
between the molecular dynamics simulations and the experiment 
was not planned (the number of particles and the size of the system are  too small
), the simulations turns out to be in a qualitative agreement with 
the experiment. 
We have obtained a 
gas phase for high field levels  and 
spontaneous formation of 
clusters  and coarsening for lower filed level, see Fig. \ref{fig5}. 
However, 
detailed  
simulations with much large number of particles and more realistic account
for  sphere polarization effects  are necessary to achieve quantitative 
agreement with experiment.

We have reported a hysteretic first-order phase
transition accompanied by
coarsening behavior in the electrostatically-driven conducting granular
medium.
The origin of coarsening dynamics and
hysteresis is due to a  screening of the electric field
in dense particle clusters. Our results show surprisingly high sensitivity
of the phase boundaries to surface coating of the grains due to humidity.
Simplified molecular dynamics simulations with conducting particles demonstrate
qualitative agreement with
the experiment: existence of two critical fields and
coarsening.

We thank Sid Nagel, Leo  Kadanoff, Thomas Witten, 
Lorenz Kramer and Baruch Meerson
for useful discussions. This research is supported by
US DOE, grant W-31-109-ENG-38,  and by NSF, STCS \#DMR91-20000.

\begin{figure}[h]
%\centerline{  \psfig{figure=figs/Fig1_bw.ps,height=2.1in}}
\caption{The electrostatic cell}
\label{Fig1}
\end{figure}

\begin{figure}
%\centerline{ \psfig{figure=figs/fig2.ps,height=2.75in}}
\caption{
The phase diagram in 
$f,E_0$ plane. 
Lines with diamonds show $E_{1,2}$ in open cell, 
dashed (dash-dotted) lines show   $E_2$ ($E_1$) 
for the evacuated cell. In {\it gas only}  
part  clusters do not form, in {\it coarsening} 
part clusters and gas coexist.  
}
\label{Fig2}
\end{figure}
\begin{figure}
%\centerline{ \psfig{figure=figs/Fig3a.eps,height=3.062in}}
%\centerline{ \psfig{figure=figs/Fig3b.eps,height=1.05in}
% \psfig{figure=figs/Fig3c.eps,height=1.05in}}
\caption{
(a) Close-up image of a single cluster. 
The larger  particle near the center  ({\it arrow}) 
 of the cluster served 
as the nucleation point. 
Top view of the cell for   (b) $t=10$ sec, and
(c) $t=10910 $ sec for $f=40$ Hz  and $E=5.25$ kV/cm.   
}
\label{Fig3}
\end{figure}

\begin{figure}
%\centerline{ \psfig{figure=figs/Fig4.eps,height=2.7in}}
\caption{
$\langle R \rangle $ vs $t^{1/2}$ for 
open cell ($\circ$), evacuated cell ($\bullet$) 
and $1/N$  for open cell ($\diamond$). 
The conditions were $E=5.25$ kV/cm and $f=40$ Hz for 
open cell and $E=4$ kV/cm and $f=8$ Hz for evacuated cell. 
Dashed lines show linear fit to these curves. 
Inset: $N$ vs $t$ for open cell ($\circ$). 
Dashed lines show power-law fits 
$N \sim  t^{-1/2} $ and $N \sim 1/ t $ correspondingly. }
\label{Fig4}
\end{figure}
\begin{figure}
%\centerline{ \psfig{figure=figs/fig5.ps,width=3.5in} }
\caption{
Top view of simulation domain. Size of the domain is $200\times200$ particle 
diameters, number of particles is  3000,  value of the applied field is  $E=1.1E_1$.  
The configurations are shown 
for $t=1$sec (a) and $t=10$sec (b). Black bullets   correspond to immobile
particles in the clusters,  open circles correspond to      
flying particles. 
}
\label{fig5}
\end{figure}
\end{document}